\documentclass[twocolumn,notitlepage,prb,color,superscriptaddress,psfig,showpacs,amsmath,amssymb,nobibnotes,longbibliography]{revtex4-2}

\usepackage[colorlinks=true,citecolor=blue]{hyperref}%

\usepackage{epsf}
\usepackage{graphicx}
\usepackage{amssymb}
\usepackage{color}
\usepackage{float}
\usepackage{epstopdf}
\usepackage{natbib}
\usepackage{amsmath}
\usepackage{mathrsfs}

\usepackage{xspace}

\usepackage{soul}
\setstcolor{red}

\newcommand{\al}[1]{\textcolor{black}{#1}}
\def\MnOTF{MnBi$_{\text{2}}$Te$_{\text{4}}$}
\def\MnOFS{MnBi$_{\text{4}}$Te$_{\text{7}}$}

\newcommand{\MBT}{(MnBi\textsubscript{2}Te\textsubscript{4})(Bi\textsubscript{2}Te\textsubscript{3})\textsubscript{n}}
\newcommand{\HIIc}{$H \parallel c$}
\newcommand{\HIIab}{$H \parallel ab$}

\definecolor{darkgreen}{rgb}{0, 0.4, 0}
\newcommand{\vk}[1]{{\color{black} { #1}}}

\newcommand{\beginappendix}{%
	\setcounter{figure}{0}
	\renewcommand{\thefigure}{A\arabic{figure}}%
}

\begin{document}
	
\title{Magnetic field tuning of the spin dynamics in the magnetic topological insulators \MBT }

\author{A.~Alfonsov}
\thanks{These authors contributed equally to this work.}
\affiliation{Leibniz IFW Dresden, D-01069 Dresden, Germany}
\author{K.~Mehlawat}
\thanks{These authors contributed equally to this work.}
\affiliation{Leibniz IFW Dresden, D-01069 Dresden, Germany}
\affiliation{Institute for Solid State and Materials Physics and W{\"u}rzburg-Dresden Cluster of Excellence ct.qmat, TU Dresden, D-01062 Dresden, Germany}
\author{A.~Zeugner}
\affiliation{H.C. Starck Tungsten GmbH, Im Schleeke 78 – 91, 38642 Goslar, Germany}
\affiliation{Former: Faculty of Chemistry and Food Chemistry, TU Dresden, D-01062 Dresden, Germany}
\author{A.~Isaeva}
\affiliation{Leibniz IFW Dresden, D-01069 Dresden, Germany}
\affiliation{Van der Waals-Zeeman Institute, Institute of Physics, University of Amsterdam, Science Park 904, 1098 XH Amsterdam, The Netherlands}
\author{B.~B\"uchner}
\affiliation{Leibniz IFW Dresden, D-01069 Dresden, Germany}
\affiliation{Institute for Solid State and Materials Physics and W{\"u}rzburg-Dresden Cluster of Excellence ct.qmat, TU Dresden, D-01062 Dresden, Germany}
\author{V.~Kataev}
\affiliation{Leibniz IFW Dresden, D-01069 Dresden, Germany}

\date{\today}
	
\begin{abstract}
	
We report a high frequency/high magnetic field electron spin resonance (HF-ESR) spectroscopy study in the sub-THz frequency domain of the two representatives of the family of magnetic topological insulators \MBT\ with $n = 0$ and 1. The HF-ESR measurements in the magnetically ordered state at a low temperature of $T = 4$\,K combined with the calculations of the  resonance modes showed that the spin dynamics in \MnOFS\ is typical for an anisotropic easy-axis type ferromagnet whereas \MnOTF\ demonstrates excitations of an anisotropic easy-axis type antiferromagnet. 
However, by applying the field stronger than a threshold value $\sim 6$\,T we observed in \MnOTF\ a crossover from the antiferromagnetic (AFM) resonance modes to the ferromagnetic (FM) modes which properties are very similar to the FM response of \MnOFS. We attribute this remarkably unusual effect unexpected for a canonical easy-axis antiferromagnet, which, additionally, can be accurately reproduced by numerical calculations of the excitation modes, to the closeness of the strength of the AFM exchange and magnetic anisotropy energies which appears to be a very specific feature of this compound. Our experimental data evidences that the spin dynamics of the magnetic building blocks of these compounds, the Mn-based septuple layers (SLs), is inherently ferromagnetic featuring persisting short-range FM correlations far above the magnetic ordering temperature as soon as the SLs get decoupled either by introducing a nonmagnetic quintuple interlayer, as in \MnOFS, or by applying a moderate magnetic field, as in \MnOTF, which may have an effect on the surface topological band structure of these compounds.

\end{abstract}
	
	\maketitle

\section{Introduction}

The recent discovery of magnetic topological insulators (TIs) where the long-range magnetic order coexists with the topologically nontrivial electronic band structure has provided a rich experimental basis for the realization of new emerging physical effects \cite{Chen2010,Xu2012,Armitage2019,Tokura2019}. In particular,  the breaking of time reversal symmetry in the magnetically ordered state offers great opportunities to explore such novel phenomena as the quantum anomalous Hall effect, the topological magnetoelectric effect, and majorana fermions \cite{Tokura2019,Chang2013,He2017,Deng2020}.  
The first representatives of this new class of materials featuring the distinguishing surface property of a TI, the surface Dirac cones, and the long-range magnetically ordered regular spin lattice in the bulk are the van der Waals layered compounds \MnOTF\, and \MnOFS\ belonging to the \MBT \vk{( n = 0 -- 3)} family \cite{Otrokov2019,gong2019experimental,Lieaaw5685,Vidal2019,hu2020van}. 

Crystallographically, \MnOTF\ is similar to the well known nonmagnetic TI Bi$_{2}$Te$_{3}$ which unit cell consists of the Te-Bi-Te-Bi-Te quintuple layers (QLs) with the difference that \vk{two additional Mn-Te atomic planes are inserted into the structural block. } The resulting Te-Bi-Te-Mn-Te-Bi-Te septuple layers (SLs) are stacked along the \vk{crystallographic} $c$~axis and held together by weak van der Waals forces \cite{Zeugner2019,PhysRevMaterials.3.064202}. In \MnOFS, the magnetic SLs are separated from each other by additional nonmagnetic QLs \vk{forming a joint SL-QL-SL-QL-SL... stack \cite{Aliev2019,Wueaax9989,souchay2019layered,Vidal2019}.}

The magnetic lattice in both compounds is built of the localized magnetic moments associated with the Mn$^{2+}$ ions which form an A-type uniaxial antiferromagnetically (AFM) ordered structure below the transition temperature $T_{\rm N} = 24$\,K and 13\,K for \MnOTF\ and \MnOFS, respectively \cite{Lieaaw5685,Otrokov2019,gong2019experimental,Lee2019,PhysRevMaterials.3.064202,Zhang2019,chen2019intrinsic,souchay2019layered,Vidal2019,hu2020van,Wueaax9989}. The AFM ordered ground state is realized \vk{due to the 
AFM interlayer coupling of the ferromagnetic (FM)}
 Mn-containing SLs. Its strength gradually decreases with \vk{an} increasing number $n$ of the QLs in the \MBT\ family. Consequently, the AFM order weakens such that an FM state could be stabilized in \MnOFS\ below 5\,K \vk{according to Refs.~\cite{Vidal2019,hu2020van} whereas it was argued in Ref.~\cite{PhysRevLett.124.197201} that the low-temperature state of \MnOFS\ is a bistable meta\-magnetic state.}  The family members with $n > 2$ behave as ferromagnets \cite{Wu2020,Klimovskikh2020}.

The properties of the magnetically ordered states in the \MBT\ family of magnetic TIs seem to be well established by now. However the spin dynamics and magnetic excitations were only scarcely addressed although their possible significance for the topological phenomena was already pointed out recently \cite{Otrokov2019,Vidal2019,Lee2019,Li2020,Alfonsov2021}. 

In this work, we study the low-energy spin dynamics in the magnetically ordered state of the single crystalline samples of \MnOTF\ and \MnOFS\ with high-field electron spin resonance (HF-ESR) spectroscopy. While \MnOFS\ exhibits magnetic excitations typical for a uniaxial ferromagnet in the entire field range of the study up to 14\,T, we observe in \MnOTF\ a remarkable qualitative change of the spin dynamics from the AFM-type in fields smaller than $\sim 6$\,T to the FM-type at higher fields. This is at odds with the expected behavior of a conventional uniaxial antiferromagnet where the excitations remain of the AFM-type above the field of the spin-flop transition and even up to the field of the forced full polarization of the magnetic sublattices. We explain this effect as being due to the comparative scales of the interplanar AFM exchange interaction energy and of the magnetic anisotropy energy of the individual septuple layers in \MnOTF. In this rare situation a moderate magnetic field promotes FM resonance of the individual SLs similar to the FM response of \MnOFS\ where the inerplanar exchange is significantly weakened. We demonstrate that this dynamic, intrinsically FM response of the magnetic layers extends in \MnOFS\ up to temperatures significantly higher than $T_{\rm N}$ suggesting that the underlying FM spin-spin correlations are prominent also in the paramagnetic state. Also in \MnOTF, in the field regime above $\sim 6$\,T where the spin wave excitations become ferromagnetic, quasi-static FM correlations are manifesting above $T_{\rm N}$. On the short timescale they can generate instantaneous internal fields possibly influencing the properties of the surface topological electronic states in either of these two materials. 

Finally, we observe in \MnOTF\ peculiar nonresonant step-like changes of the microwave absorption in the ordered state whose field position seems to follow the magnetic order parameter with a critical exponent somewhat different from what we find for the $T$-dependence of the zero-field excitation gap.

\section{Methods}

\subsection{Experimental details}
\label{experimental}

\vk{High-quality single crystals of \MnOTF\ and \MnOFS\  were grown by cooling stoichiometric mixtures of MnTe and Bi$_2$Te$_3$, as described in Refs.~\cite{Vidal2019,Zeugner2019}. Platelet-like crystals were mechanically extracted from a quenched melt and characterized by x-ray diffraction and energy-dispersive x-ray spectroscopy as sub-stoichiometric  ${\rm (Mn_{1-x}Bi_{2+2/3x}Te_4)(Bi_2Te_3)_n}$, $n = 0, 1$, $x \sim 0.2$.} The electron spin- (ESR), ferromagnetic- (FMR) and antiferromagnetic resonance (AFMR) measurements were carried out using a homemade high-field/high-frequency ESR (HF-ESR) spectrometer, equipped with a vector network analyzer (PNA-X from Keysight Technologies) for generation and detection of microwaves in a frequency range from 30 to 330\,GHz and a 16\,T superconducting magnet system (Oxford Instruments) with a variable temperature insert (VTI). A transmission probe head utilizing the oversized waveguides to guide the microwave radiation to the sample was inserted in the VTI enabling variation of the sample temperature in a range  3 -- 300\,K. 

For the sake of simplicity in the following  all measured spectra in both magnetically ordered and paramagnetic states of \MnOTF\ and \MnOFS\  will be referred to as HF-ESR spectra.

Most of the measurements were carried out in the field-sweep mode at a chosen constant microwave radiation frequency $\nu$. The signal at the detector $S_D(H)$ was recorded as a function of the applied magnetic field $H$ by continuously sweeping the field from 0 to 16\,T and then back to 0\,T. 
These experiments were complemented by measurements in the frequency-sweep mode 
at different constant magnetic field values. In order to eliminate the linear non-resonant background as well as a strong dependence of the absolute level of $S_D$ on the microwave frequency due to the complex impedance of the oversized waveguides, the ratio $S_D(\nu, H)/S_D(\nu, H-\delta H)$ was recorded at the discrete values of the magnetic field $H$ at intervals $\delta H = 0.05 - 0.2$ T. The so acquired two-dimensional $\nu - H$ data set for both
incrementally increasing ($0 \rightarrow 16$\,T) and decreasing ($16 \rightarrow 0$\,T) field values was analyzed to determine the frequency dependence of the resonance fields of the detected signals. To verify the correctness of this analysis the results were compared with the resonance positions of the signals in the spectra measured in the field-sweep mode at selected fixed frequencies $\nu$ (see Appendix, Fig.~\ref{2D}).

\subsection{Computational approach}
\label{computation}

To analyze the measured $\nu\,vs.\,H$ dependences of the FMR and AFMR modes in the magnetically ordered state of \MnOTF\ and \MnOFS,  that correspond to the collective excitations of the spin lattice (spin waves), we employed a linear spin wave theory with the second quantization formalism \cite{Turov,Holstein1940}. The phenomenological Hamiltonian for the two-sublattice spin system, as realized in the studied compounds, has the following form: 

\begin{align}
\label{Hamil}
\mathscr{H} = \ & A_{\rm inter} \frac{(\boldsymbol{M_1 M_2})}{M_0^2} + \frac{K}{2} \frac{{M_{1_z}}^2 + {M_{2_z}}^2}{M_0^2} \nonumber \\
- & (\boldsymbol{H M_1}) - (\boldsymbol{H M_2}) + 2 \pi ({M_{1_z}} + {M_{2_z}})^2\, .
\end{align} 

Here the first term represents the exchange interaction between the septuple layers with respective sublattice magnetizations $\boldsymbol{M_1 \ \text{and} \ M_2}$. $A_{\rm inter}$ is the interlayer exchange constant, and $\boldsymbol{M_1 \ \text{and} \ M_2}$ are the sums of all magnetic moments in the odd and even layers, respectively, such that $\boldsymbol{M_1}^2 = \boldsymbol{M_2}^2 = (M_0)^2 = (M_s/2)^2$, with $M_s^2$ being the square of the saturation magnetization. The second term in Eq.~(\ref{Hamil}) is the magnetocrystalline anisotropy given by the anisotropy constant $K$. It is divided by a factor of two in order to have a proper transformation of Eq.~(\ref{Hamil}) into a common form of a Hamiltonian for a ferromagnet if $A_{\rm inter} = 0$ and $\boldsymbol{M_1} = \boldsymbol{M_2}$. The third and fourth terms are the Zeeman interactions for both sublattice magnetizations. The last term stands for the shape anisotropy given by the flake-like shape of the samples. 

In order to calculate the energies of the spin waves with the wave vector $q = 0$ which are accessible by HF-ESR spectroscopy, we introduce the magnetization operators using the Holstein–Primakoff transformation. In the local coordinate frame for the respective sublattice magnetization they take the following form \cite{Holstein1940}:
\begin{align}
	\label{HP}
	&M_{j_x} = \sqrt{\frac{g \mu _{\rm B} M_0}{2}} (b_j + b_j^\dagger); \ \ M_{j_y} = i \sqrt{\frac{g \mu _{\rm B} M_0}{2}} (b_j - b_j^\dagger); \nonumber \\
	&M_{j_z} = M_0 - g \mu _{\rm B} b_j^{\dagger} b_{j}; \ \ \ j = 1,2\ .
\end{align}

Note, that here the creation ($b_j^{\dagger}$) and annihilation ($b_j$) operators are momentum independent since we are interested only in the spin waves with $q = 0$. Substitution of Eq.~(\ref{HP}) in Eq.~(\ref{Hamil}) yields the Hamiltonian in the form
$\mathscr{H} = \mathscr{H}_0 + \mathscr{H}_1 + \mathscr{H}_2 + ({ higher\,order\,terms})$. Here the term $\mathscr{H}_0$ independent of $b_j^{\dagger}$ and $b_j$ operators represents the free energy density, whose minimization yields the ground state with a certain alignment of the sublattice magnetizations $\boldsymbol{M_1 \ \text{and} \ M_2}$. $\mathscr{H}_1$ turns to $0$ at the minimum of $\mathscr{H}_0$ \cite{Turov}. Finally, $\mathscr{H}_2$, which describes the linear spin waves with $q = 0$ in the system, can be written in the form: 

\begin{align}
\label{H2}
\mathscr{H}_2 = \sum_{i,j}(\alpha_{i,j} b_i^\dagger b_j + \alpha^*_{i,j} b_i b_j^\dagger + \beta_{i,j} b_i^\dagger b_j^\dagger + \beta_{i,j}^* b_i b_j)\,.
\end{align} 

Here the coefficients $\alpha_{i,j} , \ \alpha^*_{i,j} , \ \beta_{i,j} , \ \beta_{i,j}^*$ depend on the parameters of the Hamiltonian 
Eq.~(\ref{Hamil})
and the indexes $i,\ j$ run through all the sublattice numbers, i.e. 1 and 2 in the case of a two-sublattice system. Diagonalization of Eq.~(\ref{H2}) yields the spin waves energies $\epsilon_{\rm sw}$ and the corresponding frequencies $\omega_{\rm sw} = \epsilon_{\rm sw}/\hbar$, which can be compared with the experimentally measured values. For the case of FMR, these frequencies as a function of the magnetic field were calculated in the analytical form, see Eqs.~(\ref{eq1}) and (\ref{eq2}). The AFMR modes were calculated numerically using the method suggested by C. Tsallis \cite{Tsallis1978}. 

\section{Results and discussion}

\subsection{HF-ESR spectroscopy on M\lowercase{n}B\lowercase{i}$_4$T\lowercase{e}$_7$}
\label{ESR147}

	\begin{figure*}[ht]
	\centering
	\includegraphics[clip,width=2 \columnwidth]{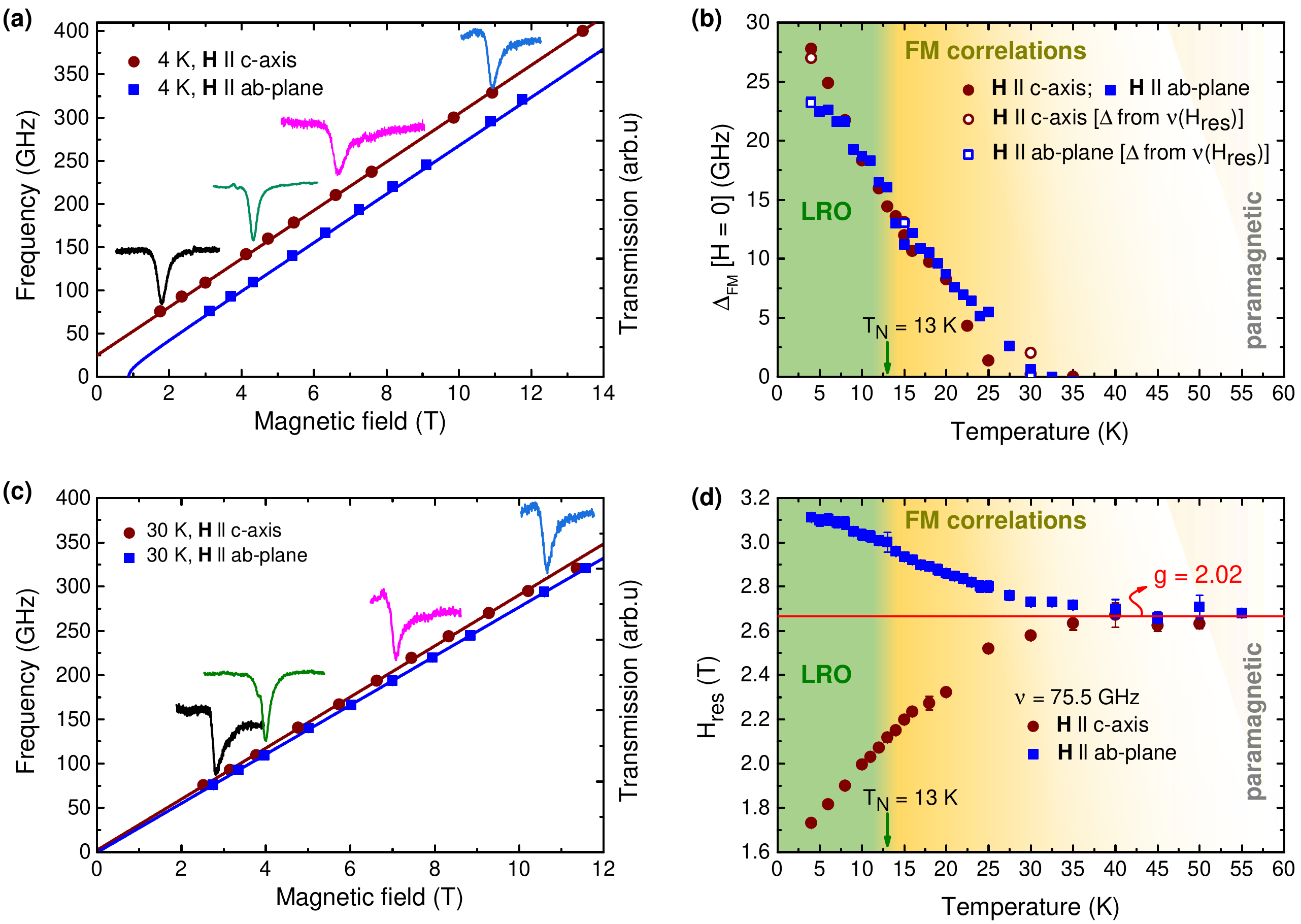}
	\caption{\MnOFS\,: Frequency versus field $\nu(H)$ diagram of the resonance  modes (symbols) at $T = 4$\,K (a) and $T = 30$\,K (c) for the external magnetic field applied parallel and perpendicular to the $c$-axis (left vertical scale) and selected HF-ESR spectra at various excitation frequencies (right vertical scale). Solid lines depict the calculated $\nu(H)$ resonance branches (see the text); (b) The excitation gap $\Delta_{\rm FM}(H = 0)$ as a function of temperature. 
Open symbols correspond to $\Delta_{\rm FM}(H = 0)$ evaluated from the $\nu(H)$ dependence measured at selected fixed temperatures for ${\bf H}\parallel c$-axis and ${\bf H}\parallel ab$-plane. Closed symbols are the gap estimates from the $T$-dependence of the resonance field\al{s measured} at \al{two} fixed frequenc\al{ies} $\nu = 75.5$\,GHz and $\nu = 219.5$\,GHz for the same two field geometries;
%
(d)  Resonance fields as a function of temperature measured at $\nu = 75.5$ GHz for ${\bf H}\parallel c$-axis and ${\bf H}\parallel ab$-plane. Horizontal solid line indicates the paramagnetic resonance field corresponding to the $g$-factor $g = 2.02$.  } 
	\label{fig:FMR-Mn147}
\end{figure*}

\subsubsection{HF-ESR modes  in the ordered state}

\paragraph*{\underline{Experimental data}:}

To obtain the frequency-magnetic field $\nu(H)$ diagram of the resonance modes in the ordered state of \MnOFS\ the HF-ESR spectra were measured at $T = 4$ K $<$ $T_{\rm N} = 13$\,K \cite{Vidal2019} for the external magnetic field applied in the directions ${\bf H}\parallel c$-axis and ${\bf H}\parallel ab$-plane over the frequency range 75 -- 330\,GHz. The ESR spectra consist of a single resonance line for both magnetic field configurations. The resonance field $H_{\rm res}$ at a given excitation frequency was determined by fitting the spectra with a Lorentzian line profile. The so obtained resonance branches $\nu(H_{\rm res})$ are plotted
in Fig.~\ref{fig:FMR-Mn147}(a) (left vertical scale) together with representative HF-ESR spectra (right vertical scale). 

\paragraph*{\underline{Exchange and magnetic anisotropy parameters}:}

The $\nu(H_{\rm res})$ dependence for the two field geometries was analyzed with the theoretical model introduced in Sect.~\ref{computation} which has enabled to classify the measured signals as FMR modes of a ferromagnet with uniaxial anisotropy. In this case the analytical solution is available yielding the resonance frequencies \cite{Turov}  
\begin{equation}
	\label{eq1}
	2\pi\nu_{\rm c} = \gamma({H + H_{A_{\rm T}}})
\end{equation}
and 
\begin{equation}
	\label{eq2}
	2\pi\nu_{\rm ab} = \gamma \sqrt{ H(H - H_{A_{\rm T}}) }, \, \, (H > H_{A_{\rm T}})
\end{equation} 
for ${\bf H}\parallel c$-axis and ${\bf H}\parallel ab$-plane, respectively.  
Here $\gamma = g \mu _{\rm B}/\hbar$ is the  gyromagnetic ratio with $g$ being the effective $g$-factor in the ordered state, $\mu_{\rm B}$ is the Bohr magneton and $\hbar$ is the reduced Planck constant. $H_{A_{\rm T}} = H_{\rm A} + H_{\rm D}$ is the total magnetic anisotropy  field comprising the demagnetization field for an infinite FM plane $H_{\rm D} = 4 \pi M_{\rm s}$ \cite{Blundell2001} and the magnetocrystalline anisotropy field $H_{\rm A} = 2K/M_{\rm s}$. The demagnetization field corresponds to the thin-plate shape of the \MnOFS\, single crystal used in these HF-ESR measurements. For such sample shape this field can be straightforwardly calculated yielding the value of 0.33\,T. The fit of the frequency-field dependence of the FMR using Eqs.~(\ref{eq1}) and (\ref{eq2}) provides the estimates of the total magnetic anisotropy field $H_{A_{\rm T}} = - 0.82 \pm 0.02$\,T and of the effective $g$-factor of $2.02 \pm 0.003$ for the ${\bf H}\parallel ab$-plane configuration. 
For the ${\bf H}\parallel c$-axis configuration, the fit yields the values of the ferromagnetic zero field excitation gap $\Delta_{\rm FM}  \vert _{H = 0}  \equiv \Delta_{\rm FM} = \vert\gamma H_{A_{\rm T}} \vert = 27 \pm 0.6$\,GHz, corresponding to $H_{A_{\rm T}} = -0.97 \pm 0.02$ T, and the $g$-factor of $1.98 \pm 0.003$. Using the  values of $H_{A_{\rm T}}$ for both field configurations we can estimate the average total magnetic anisotropy field $H_{A_{\rm T}} = -0.9 \pm 0.08$ T, and from it the value of the intrinsic magnetocrystalline anisotropy $H_{\rm A} = -1.23 \pm 0.08$\,T. The corresponding value of the anisotropy constant $K$ for \MnOFS\ amounts then to $K_{147} = H_{\rm A} M_{\rm s}/2 = - 1.6 \pm 0.1 \times 10^{6}$\,erg/cm$^{3}$. The negative sign of $K$ signifies the easy axis magnetic anisotropy of \MnOFS.

Noting that the unit cell of \MnOFS\ contains one (nonmagnetic) QL and one (magnetic) SL the magnetic anisotropy energy (MAE) of $E_{\rm MAE} =- 0.18 \pm 0.01$\,meV per single SL can be derived from $K_{147}$ which should correspond then to the MAE of the \MnOTF\ unit cell consisting of only one septuple layer  
\footnote{In order to consistently evaluate the MAE for both \MnOFS\ and \MnOTF\ compounds, we defined the saturation magnetization for one formula unit (f.u.) in the single septuple layer. The unit volumes, therefore, are $179.732\cdot 10^{-24}$ cm\textsuperscript{3} for \MnOFS, and $177.314\cdot 10^{-24}$ cm\textsuperscript{3} for \MnOTF, respectively. Assuming the full saturation of Mn\textsuperscript{2+} (S = 5/2, L = 0, $g = 2$) moment $M_s = 5 \mu _B / \text{Mn}$, the saturation magnetization per f.u. amounts to $\sim 258$ erg/(G cm\textsuperscript{3}) for \MnOFS, and to $\sim 261.5$ erg/(G cm\textsuperscript{3}) for \MnOTF.}. 

\subsubsection{Temperature dependence of HF-ESR response}
\label{subsubsect:FMR-Tdep-Mn147}

To examine the evolution of the ferromagneticaly ordered state in \MnOFS\ upon increasing temperature the $\nu(H)$ dependence of the resonance signals was measured at different temperatures and evaluated using Eqs.~(\ref{eq1}) and (\ref{eq2}). The resulting $T$-dependence of the FM excitation gap $\Delta_{\rm FM}(T)$  is shown in Fig.~\ref{fig:FMR-Mn147}(b) and that of the resonance field $H_{\rm res}(T)$ at a selected frequency $\nu \approx 75$\,GHz is presented in Fig.~\ref{fig:FMR-Mn147}(d).
Remarkably, the gap does not close at $T_{\rm N} = 13$\,K but persists up to a much higher temperature of $\sim 30$\,K evidencing the presence of short range ferromagnetic spin correlations in the SLs of \MnOFS, static on the HF-ESR time scale (10$ ^{-10} $-10$ ^{-11} $ s) even in the paramagnetic state. The presence of quasi-static internal fields, associated with the intrinsic magnetocrystalline anisotropy and the ferromagnetic exchange that give rise to the anisotropy of $H_{\rm res}$, can be well seen in the temperature dependence of the ESR line shift shown in Fig.\ref{fig:FMR-Mn147}(\al{d}). Note, that in the purely paramagnetic state the ESR signal of Mn$^{2+}$ ($3d^5, S = 5/2, L = 0$) is nearly isotropic due to the absence of the orbital momentum of the half-filled 3$d$ shell of this ion in first order \cite{AbragamBleaney}. Indeed, at $T=30$\,K, where the zero field gap is practically closed, the resonance branches $\nu(H)$ for both field geometries obey a simple paramagnetic resonance condition $ \nu = (g\mu_{\rm B}/h)H_{\rm res}$ as shown in Fig.~\ref{fig:FMR-Mn147}(c). The fit yields the corresponding $g$-factor values $g_{\rm c} = 2.06$ and $g_{\rm ab} = 1.98$, which are very close to the spin-only value as expected for the paramagnetic spin-only moment of the Mn$^{2+}$ ion. 

The fact that the resonance parameters do not change abruptly at the ordering transition but the magnetic system in \MnOFS\ demonstrates a rather smooth, crossover-like evolution of the spin dynamics from the long-range ordered (LRO) state to an extended short-range FM correlated regime above $T_{\rm N}$ before entering the uncorrelated paramagnetic state at even higher temperatures, as pictorially indicated by shaded areas in Figs.~\ref{fig:FMR-Mn147}(b) and (d), suggests a low-dimensional (low-D) character of the spin system in \MnOFS. This kind of behavior was extensively addressed in the literature in the past, e.g., for the case of ESR in 1D spin lattices \cite{Okuda1972,Nagata1972,Oshima1976,Brockmann2012,Zeisner2017} but also in 2D-systems \cite{Benner1990,Zeisner2019,Zeisner2020}. Indeed, in \MnOFS\ the magnetic SLs are separated by nonmagnetic QLs along the $c$-axis which effectively reduces the magnetic dimensionality and weakens the interlayer antiferromagnetic coupling. Therefore, \MnOFS\ features an inherently ferromagnetic spin correlated dynamics decoupled from the fragile antiferromagnetic 3D long-range order occurring at small fields at $T_{\rm N} = 13$\,K.

\subsection{HF-ESR spectroscopy on M\lowercase{n}B\lowercase{i}$_2$T\lowercase{e}$_4$}

\subsubsection{HF-ESR $\nu(H)$ diagram  in the ordered state}
\label{subsubsect:nu-H-diag-M124}

\begin{figure}[h]
	\centering
	\includegraphics[clip,width=0.90\columnwidth]{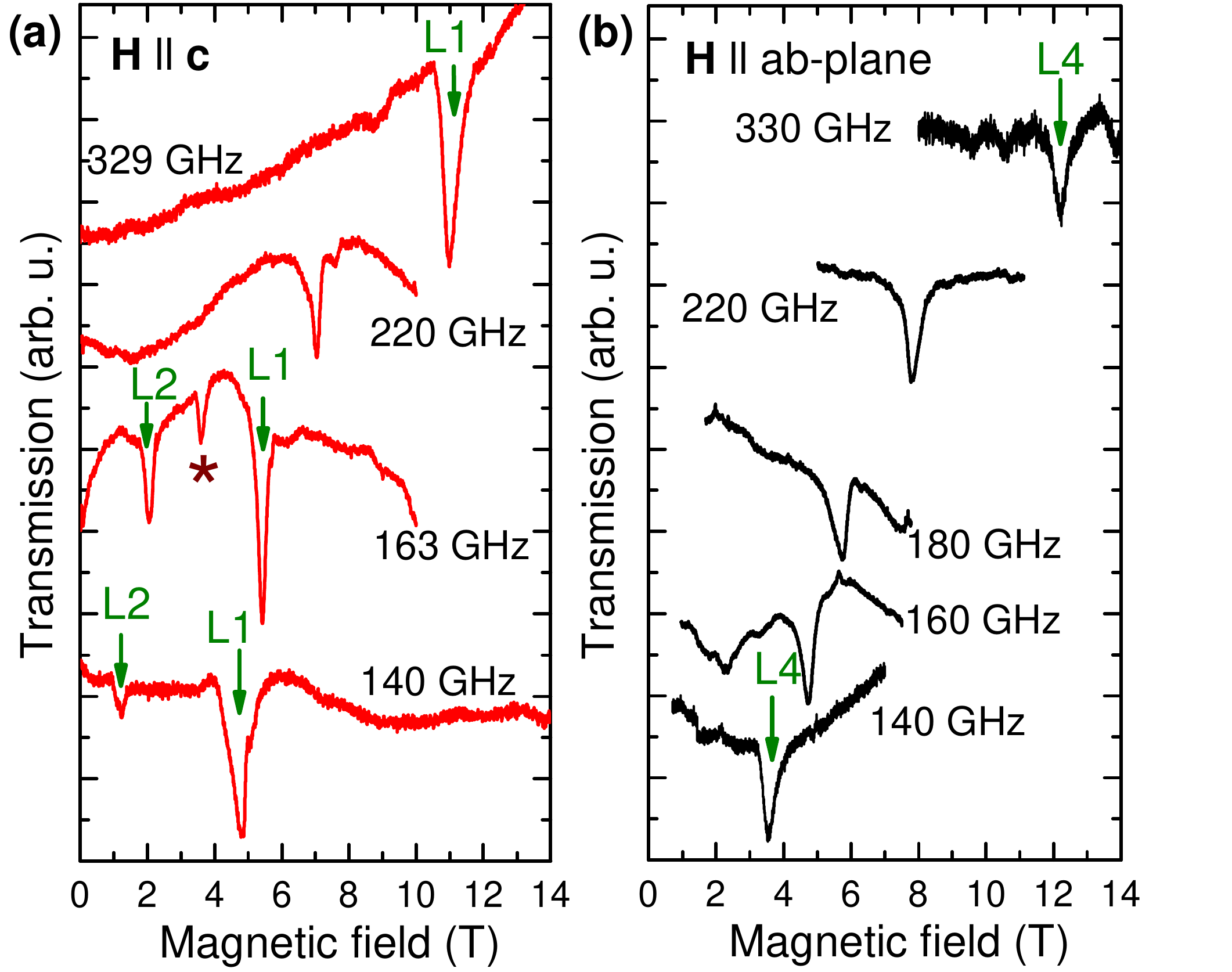}
	\caption{\MnOTF\,: HF-ESR spectra at selected excitation frequencies recorded at $T = 4$\,K for two orientations of the external magnetic field ${\bf H}\parallel c$-axis (a) and ${\bf H}\parallel ab$-plane (b). For clarity, all spectra are normalized and shifted vertically. Labels $L1$, $L2$ and $L4$ indicate the resonance modes, and $*$ marks the impurity peak.} 
	\label{fig:AFMR-Mn124}
\end{figure}

\begin{figure}[t]
	\centering
	\includegraphics[clip,width=0.90\columnwidth]{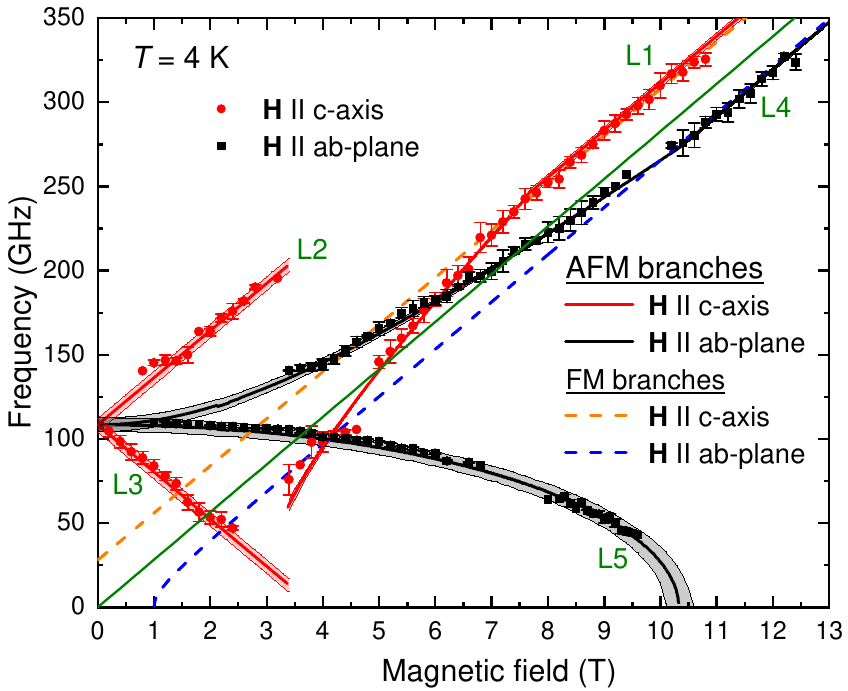}
	\caption{\MnOTF\,: Frequency versus field $\nu(H)$ diagram of the resonance  modes (symbols) at $T = 4$\,K for the external magnetic field applied parallel and perpendicular to the $c$-axis. Black and red solid curves are the results of the modeling of the resonance branches. The shaded area around the curves depicts the calculation uncertainty corresponding to the error bars of the anisotropy and exchange constants. The dashed lines correspond to the calculations of the modes of an easy-axis ferromagnet according to Eq.~(\ref{eq1}) and (\ref{eq2}). The straight green solid line represents the paramagnetic resonance condition $ \nu = (g\mu_{\rm B}/h)H_{\rm res}$.  (see the text for details)}
			\label{fig:modes-Mn124}
\end{figure}

To map the $\nu(H)$ diagram of the resonance modes in the magnetically ordered state of \MnOTF\ HF-ESR spectra
were measured at $T = 4$\,K\,$\ll T_{\rm N} = 24$\,K~\cite{Otrokov2019} in a frequency range 75 -- 330\,GHz for two orientations of the external field ${\bf H}\parallel c$-axis and ${\bf H}\parallel ab$-plane. In contrast to \MnOFS, the HF-ESR response of \MnOTF\ is more rich. Characteristic spectra recorded in the field-sweep mode at different fixed frequencies are shown in Fig.~\ref{fig:AFMR-Mn124}. Several resonance lines ascending in field with increasing the frequency are easily discernible. For ${\bf H}\parallel c$-axis, this are lines labeled L1 and L2 in Fig.~\ref{fig:AFMR-Mn124}(a) and, for ${\bf H}\parallel ab$-plane, the line labeled L4 in Fig.~\ref{fig:AFMR-Mn124}(b). For a better understanding of their frequency-magnetic field dependence complementary measurements in the frequency-sweep mode described in Sect.~\ref{experimental} were carried out. The result is shown in Fig.~\ref{2D} in Appendix. As can be seen in this Figure,  the data obtained in both measurement modes match very well and two new signals L3 and L5 descending in field with increasing the frequency could be identified in the frequency-sweep experiment. The resulting $\nu(H)$ diagram of the resonance modes is presented in Fig.~\ref{fig:modes-Mn124}.     

At fields smaller than $\sim 5$\,T the resonance branches are rather typical for a collinear two-sublattice antiferromagnet with an easy-axis anisotropy \cite{Turov}. In zero applied field the two modes are degenerate and gapped due to magnetic anisotropy of the spin sublattices. One mode corresponds to the clockwise precession of the sublattice magnetization vector ${\bf M}_1$ and the anticlockwise precession of ${\bf M}_2$ about the easy axis. For the other mode the sign of precession of ${\bf M}_1$ and ${\bf M}_2$ is reversed. The magnitude of the gap at $T = 4$\,K amounts to $\Delta_{\rm AFM}  \vert _{H = 0}  \equiv \Delta_{\rm AFM} = 108 \pm 4$\,GHz (0.45\,meV) and agrees well with the inelastic neutron scattering result \cite{Li2020}. Application of magnetic field along the easy-axis ($c$-axis) increases the frequency of one mode (branch L2) and decreases the frequency of the other mode (branch L3). The corresponding branches progress linearly in field until the spin-flop field $H_{\rm sf}$ at 3.5\,T \cite{Otrokov2019} is reached where the two magnetization vectors turn perpendicular to the applied field. At $H > H_{\rm sf}$ both magnetization vectors precess jointly giving rise to branch L1. For the magnetic field applied perpendicular to the easy-axis there are two oscillating modes, one ascending (L4) and one descending (L5) \cite{Turov}. The latter one softens to zero at the saturation field. 

Remarkably, the resonance modes  in \MnOTF\  experience a drastic transformation at fields above the spin flip transition. Branches L1 and L4 cross at $\sim 6$\,T and progress practically linear in parallel at fields above $\sim 9$\,T. This is not expected for a canonical easy-axis antiferromagnet where these two branches asymptotically approach the paramagnetic line $\nu = h^{-1}g\mu_{\rm B}H$ (green straight line in Fig.~\ref{fig:modes-Mn124}) from above and below, respectively, and never cross \cite{Turov,Rezende2020}.  Moreover, the behavior of branches L1 and L4 at high fields appear to be very similar to the ferromagnetic modes in \MnOFS\ [Fig.~\ref{fig:FMR-Mn147}(a)].

\subsubsection{Analysis and discussion of magnetic excitations}
\label{subsubsect:analysis-nu-H-diag-M124}

In order to understand such an unusual qualitative change of the spin dynamics in \MnOTF\ from the antiferromagnetic at low fields to the ferromagnetic at higher fields we performed  numerical calculations of the spin waves energies for various values of the interlayer exchange constant $A_{\rm inter}$ and the magnetocrystalline constant $K_{124}$ in Eq.~(\ref{Hamil}). The best model curves excellently fitting the experimental data for both magnetic field geometries are presented in Fig.~\ref{fig:modes-Mn124}. The fit yields 
$A_{\rm inter}=(5.8 \pm 0.1) \times 10^{6}$\,erg/cm$^{3}$ and $K_{124}=-(1.9 \pm 0.1) \times 10^{6}$\,erg/cm$^{3}$. These values correspond to the exchange energy of $E_{\rm inter-exch} = 0.65 \pm 0.01$\,meV and magnetic anisotropy energy $E_{\rm MAE} =-0.21 \pm 0.01$ meV per formula unit of \MnOTF\ \cite{Note1}. Notably, the latter value is close to the MAE of the single septuple layer in \MnOFS\ estimated in Sect.~\ref{ESR147}. This similarity  suggests that the magnetic anisotropy of an individual SL is its intrinsic property little affected by introduction of additional nonmagnetic QLs in the crystal lattice of the \MBT\ family. 

The obtained estimates of $E_{\rm inter-exch}$ and $E_{\rm MAE}$  give the clue to understand an unexpected transformation of the spin dynamics in \MnOTF\ as a function of magnetic field. The energy scale of these two parameters appears similar, whereas in most of the transition-oxide based antiferromagnets to which the canonical theory \cite{Turov} applies one finds typically $E_{\rm inter-exch} \gg E_{\rm MAE}$.  The dominance of the antiferromagnetic exchange over magnetic anisotropy maintains the AFM-type of excitations up to the fields of the full polarization of the sublattices. In contrast, in \MnOTF\  the interplane exchange over the van der Walls gap is generally weaker than in the covalently-bonded solids. Thus, essentially ferromagnetic SLs with magnetocrystalline anistotropy comparable in strength with $E_{\rm inter-exch}$ can be effectively decoupled by the applied magnetic field and demonstrate their inherent ferromagnetic dynamics as in \MnOFS. This crossover is accurately reproduced by the numerical solution of Eq.~(\ref{H2}) as demonstrated in Fig.~\ref{fig:modes-Mn124}.

Additionally, these calculations yielded the static magnetization $M(H)$ curves for two directions of the 
magnetic field (see Sect.~\ref{appendix}, Fig.~\ref{calc_magn}), which enable to estimate the values of the characteristic magnetic fields, such as the spin-flop field $H_{\rm sf}^{\text{\HIIc}} \approx 3.4$\,T, and the saturation fields $H_{\rm S}^{\text{\HIIc}} \approx 7.6$\,T and  $H_{\rm S}^{\text{\HIIab}} \approx 10.3$\,T. These values are in a reasonable agreement with the results of the magnetization measurements on \MnOTF\ \cite{Otrokov2019,Li2020} which additionally confirms the validity of the present analysis of the spin dynamics in \MnOTF. 
	
\subsubsection{Temperature evolution of the spin dynamics}	
	
\begin{figure}[t]
	\centering
	\includegraphics[clip,width=0.90\columnwidth]{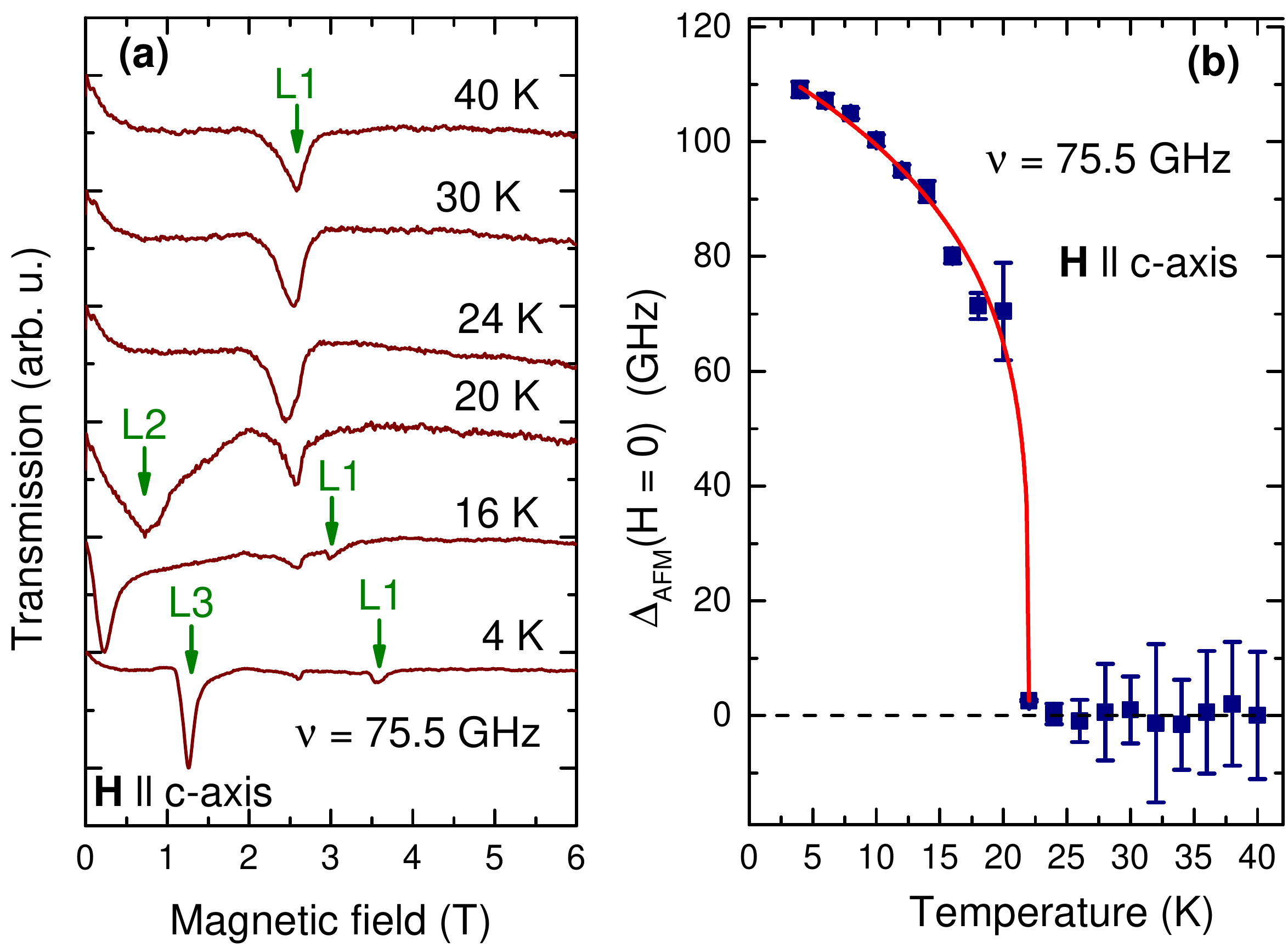}
	\caption{\MnOTF\,: (a) HF-ESR spectra at selected temperatures recorded at  $\nu  =76$\,GHz for ${\bf H}\parallel c$-axis. Labels L1 and L3 indicate the resonance modes. The spectra are normalized and shifted vertically for clarity; (b) AFM zero field gap as a function of temperature (symbols). The solid line is a fit using 
	Eq.~(\ref{eq10}).} 
	\label{fig:gap-Mn124}
\end{figure}

\paragraph*{\underline{Excitation gap}:}
The AFM zero field gap for HF-ESR excitations, i.e. the magnon gap at the wave vector $q=0$,  $\Delta_{\rm AFM} \vert _{T = 4\,{\rm K}} = 108 \pm 4$\,GHz (0.45\,meV) was straightforwardly determined as the intercept of branches L2 and L3 with the frequency axis (Fig.~\ref{fig:modes-Mn124}). The temperature dependence of $\Delta_{\rm AFM}$ was obtained in a similar way by measuring the HF-ESR spectra \al{in the field range below SF} at different temperatures and excitation frequencies for ${\bf H}\parallel c$-axis.  
As an example, the temperature dependence of the spectra measured at $\nu = 75.5$\,GHz is presented in Fig.\ref{fig:gap-Mn124}(a).
The modes L2, L3 and L1 observed below $T_{\rm N} = 24$\,K gradually shift with increasing temperature until they merge into a single line, whose resonance field is temperature independent above $T_{\rm N}$. The resonance fields $H_{\rm res}$ of modes L2 and L3 at different frequencies $\nu_{\rm c}$ were fitted with the linear expression $ \nu_{\rm c} = \Delta_{\rm AFM}(T) + (g \mu _{\rm B}/h)H_{\rm res}$. The resulting $T$-dependence of the zero field gap $\Delta_{\rm AFM}(T)$ is plotted in Fig.~\ref{fig:gap-Mn124}(b).

Remarkably, the gap closes completely at the AFM phase transition temperature $T_{\rm N} = 24$\,K of \MnOTF. This is in a stark contrast to the FM gap in \MnOFS\ which persists on the timescale of HF-ESR at $T \gg T_{\rm N}$ [Fig.~\ref{fig:FMR-Mn147}(b)]. This hints to a more 3D character of \al{(AFM)} magnetic order in \MnOTF\ as compared to \MnOFS\ manifesting pronounced low-D FM correlations above $T_{\rm N}$, as discussed in Sect.~\ref{subsubsect:FMR-Tdep-Mn147}.

One can tentatively assume that the $\Delta_{\rm AFM}(T)$ follows approximately the order parameter of the AFM phase of \MnOTF\ and fit it to the power law function

\begin{equation}
	\label{eq10}
	\Delta_{\rm AFM}(T) \propto [1-(T/T_{\rm N})]^{\beta} \, .
\end{equation}

The fit shown in Fig.~\ref{fig:gap-Mn124}(b) yields $T_{\rm N} = 22  \pm 0.1$\,K and the critical exponent  $\beta = 0.239 \pm 0.01$. This value of $\beta$ is significantly smaller than 0.367 for the 3D Heisenberg AFM model and lies in between 0.326 and 0.125, the values for the 3D and 2D Ising models, respectively \cite{Blundell2001,Collins1989,Jongh1990}. A significant reduction of $\beta$ towards the Ising values might be related to an appreciable easy-axis magnetic anisotropy which, in the case of $E_{\rm MAE}$ comparable with $E_{\rm inter-exch}$, as found in \MnOTF, can in fact be expected \cite{Leonel2006}. 

\begin{figure}[t]
	\centering
	\includegraphics[clip,width=0.90\columnwidth]{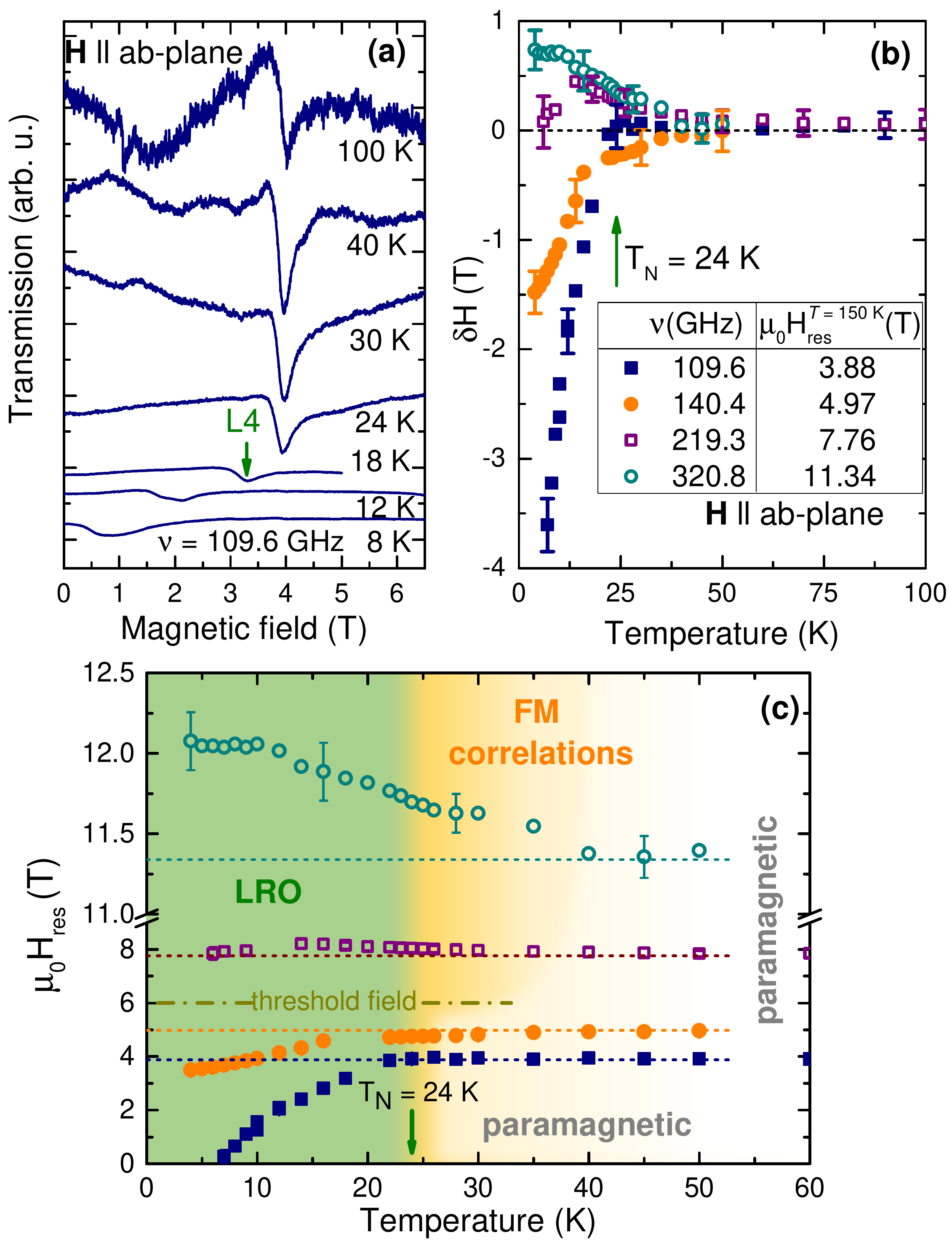}
	\caption{\MnOTF\,: (a) HF-ESR spectra (mode L4) at selected temperatures recorded at  $\nu  =109.6$\,GHz for ${\bf H}\parallel ab$-plane. The spectra are normalized and shifted vertically for clarity; (b) Resonance shift of L4 $\delta H(T) = H_{\rm res}(T) - H_{\rm res}({\rm 150\,K})$ as a function of temperature at several selected microwave frequencies for ${\bf H}\parallel ab$-plane. The corresponding resonance fields in the paramagnetic state at $T = 150$\,K  are indicated in the legend; \vk{(c) Same as (b), but instead of the relative shifts the resonance fields $H_{\rm res}$ are plotted on the absolute scale. Horizontal dashed lines represent the $T$-independent resonance fields in the paramagnetic state whose values for the given frequency $\nu$ are listed in the legend to panel (b). Horizontal dash-dot line indicates the threshold field $\sim 6$\,T for the crossover from the AFM-type to FM-type of spin excitations in \MnOTF\ with increasing the field strength. The color code depicts the long-range ordered, short range FM-correlated and paramagnetic regimes of the  the spin dynamics observed  in \MnOTF\ by HF-ESR.}
	}  
	\label{fig:shifts-Mn124}
\end{figure}

\vskip5pt
\paragraph*{\underline{Lineshifts}:}
The $T$-dependence of the resonance field of the HF-ESR signals was studied for the ${\bf H}\parallel ab$-plane field geometry by measuring mode L4 at different temperatures and frequencies $\nu$. Characteristic spectra for $\nu = 109.6$\,GHz are shown in Fig.~\ref{fig:shifts-Mn124}(a). Mode L4 gradually shifts with increasing temperature towards higher fields until reaching a constant value of the resonance field $H_{\rm res}$ at the AFM ordering temperature $T_{\rm N} = 24$\,K. The position of the line remains $T$-independent up to the highest measured temperature $T = 150$\,K and corresponds to the paramagnetic $g$-factor $g = 2.02$. As most of the measurements at other frequencies converge at high temperature to this $g$-value including the value reported in Ref.~\cite{Alfonsov2021}, the field $H_{\rm res} (T = {\rm 150\,K}) = h \nu / (g \mu_{\rm B})$ is set for each given frequency $\nu$ as the reference for an unshifted line in the purely paramagnetic uncorrelated state of \MnOTF. The $T$-dependences of the relative shifts $\delta H(T) = H_{\rm res}(T) - H_{\rm res}({\rm 150\,K})$ at selected excitation frequencies are plotted in Fig.~\ref{fig:shifts-Mn124}(b). The respective values of $H_{\rm res} ({\rm 150\,K})$ are indicated in the figure legend. The $T$-dependences of the absolute values of $H_{\rm res}$ are shown in Fig.~\ref{fig:shifts-Mn124}(c). The legend in panel (b) of this figure applies to panel (c) as well.      

As discussed in Sects.~\ref{subsubsect:nu-H-diag-M124} and \ref{subsubsect:analysis-nu-H-diag-M124}, the resonance signals in the magnetically ordered state of \MnOTF\ at small applied fields are the typical AFM excitations of an uniaxial easy-axis antiferromagnet. The negative shift of mode L4 at the smallest frequency $\nu = 109.6$\,GHz which decreases in magnitude with increasing the temperature is determined by the $T$-dependence of the AFM excitation gap $\Delta_{\rm AFM}(T)$ closing at $T_{\rm N} = 24$\,K [Fig.~\ref{fig:gap-Mn124}(b)]. At this temperature L4 reaches the $T$-independent paramagnetic position [Figs.~\ref{fig:shifts-Mn124}(b) \vk{and (c)}]. Similarly, probing L4 at a higher excitation frequency $\nu = 140.4$\,GHz and correspondingly higher magnetic field (see $\nu(H)$ dependence of branch L4 in Fig.~\ref{fig:modes-Mn124}) yields a smaller negative shift. Interestingly, it remains finite at temperature even somewhat larger than $T_{\rm N}$. At still higher probing frequencies the shift changes sign and, most remarkably, remains finite up to temperatures $T \sim 2T_{\rm N}$ [Figs.~\ref{fig:shifts-Mn124}(b) \vk{and (c)}]. The sign change of $\delta H(T)$ is related to the transformation of branch L4 from the AFM- to the FM-type at these fields where it crosses branch L1 and the paramagnetic $ \nu = (g\mu_{\rm B}/h)H_{\rm res}$  line in the $\nu(H)$ diagram shown in  Fig.~\ref{fig:modes-Mn124}.

The shift of mode L4 from its paramagnetic position in an extended temperature range above $T_{\rm N}$ in the regime when the applied field turns the resonance excitations to the FM-type signifies the persistence of the quasi-static on the HF-ESR time scale 2D ferromagnetic correlations in \MnOTF\ \al{(Fig.~\ref{fig:shifts-Mn124}(c))} in a close analogy with our observations in the case of \MnOFS\ (see Sect.~\ref{subsubsect:FMR-Tdep-Mn147}). This suggests that the ferromagnetic spin dynamics appears to be the generic property of the individual septuple layers in the \MBT\ family as soon as they become decoupled either by the magnetic field as in \MnOTF\ or by introducing a nonmagnetic interlayer (QL) as in \MnOFS. \vk{The color-coded schematic phase diagram of \MnOTF\ depicted in  Fig.~\ref{fig:shifts-Mn124}(c) indicates the three different regimes of the spin dynamics taking place in the long-range ordered state at  $T<T_{\rm N}=24$\,K, in the paramagnetic state occurring just above $T_{\rm N}$ at small fields and in the short-range ferromagnetically correlated state extending at high field up to $T \sim 2T_{\rm N}$.}

\begin{figure}[t]
	\centering
	\includegraphics[clip,width=0.90\columnwidth]{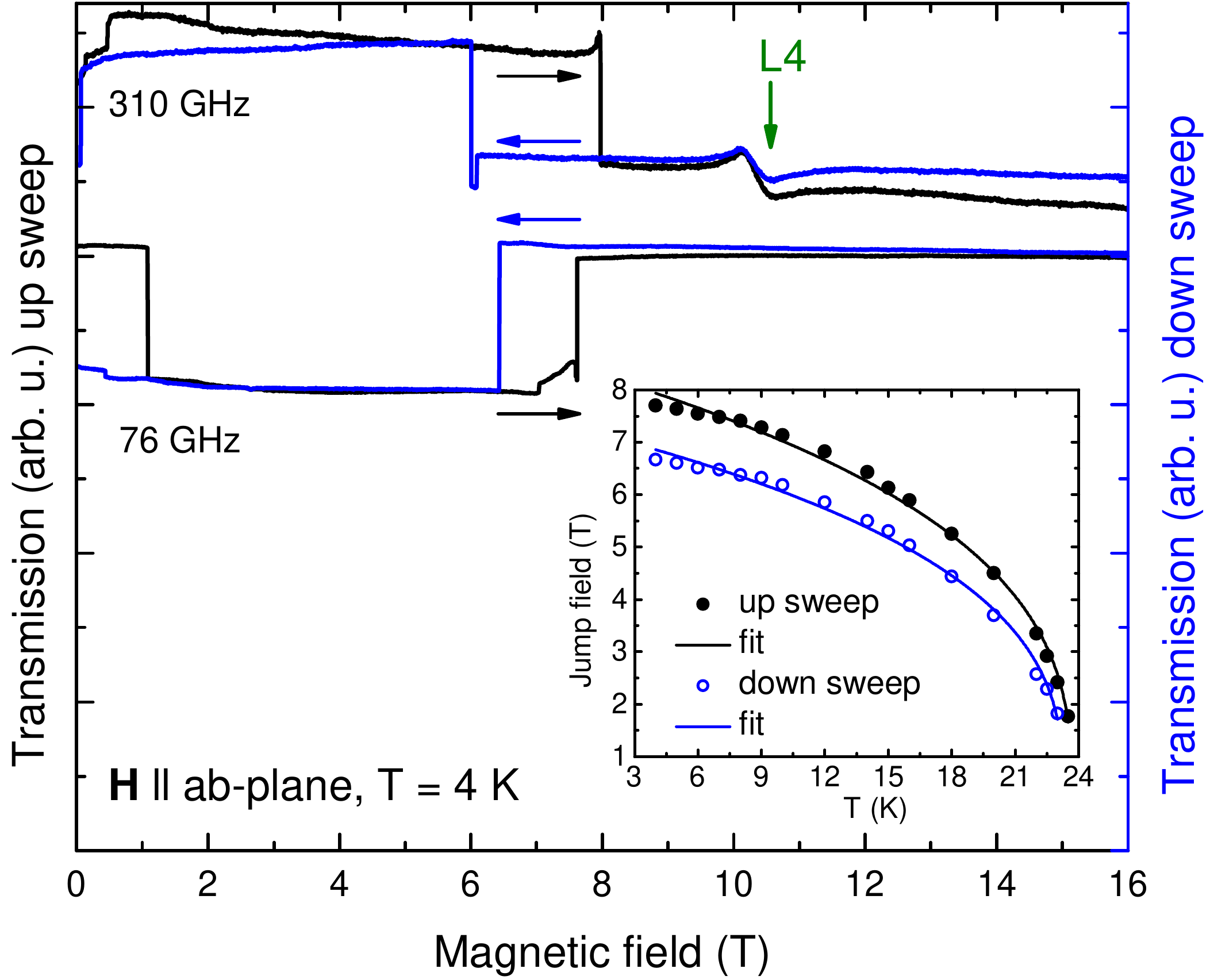}
	\caption{\MnOTF: Signal traces $S_{D}(H)$ featuring step-like changes in the range 6--8\,T recorded at $T = 4$\,K for two frequencies of 76 and 310\,GHz for the up- and down sweeps of the magnetic field ${\bf H}\parallel ab$-plane. Resonance mode L4 is indicated by the arrow. Inset shows the temperature dependence of the jump field  measured at $\nu = 109.6$\,GHz. Solid lines are the fits to the critical exponent function $\propto [1-(T/T_{\rm N})]^{\beta}$.} 
	\label{fig:jumps-Mn124}
\end{figure}

\section{Nonresonant effects}

Finally, in some of our experiments we observed a peculiar nonresonant step-like change of the microwave absorption (jumps) in the samples of \MnOTF\ while sweeping the magnetic field applied within the $ab$-plane. Examples of the signal traces $S_D(H)$ recorded at $T = 4$\,K are shown in
Fig.~\ref{fig:jumps-Mn124}. The steps occur in the field range\al{s} \al{0--1\,T and} 6--8\,T \al{and are practically} independent of frequency. They show a pronounced hysteresis upon reversing the direction of the field sweep. This is not related to a possible (small) hysteresis of the superconducting solenoid of the magnet system since the position of the resonance signal observed at $\nu = 310$\,GHz (mode L4) is practically the same for the up- and down-field sweep. The observation of the jumps in $S_D(H)$ was strongly dependent on the installation of the sample in the probe head, which points to its dependence on some uncontrollable details of  the coupling of the sample to the microwaves. But once observed, the steps occur in the same field range at a given temperature.  

The step-like changes of the microwave absorption are apparently related to the magnetism of \MnOTF. The position of the \al{high field} steps denoted as the jump field  is strongly temperature dependent. It decreases with increasing $T$ as shown in Fig.~\ref{fig:jumps-Mn124}(inset) and the steps disappear at the AFM ordering temperature $T_{\rm N}$. Fitting of this dependence with the function similar to Eq.~(\ref{eq10}) yields the ordering temperature $T_{\rm N} = 23.7 \pm 0.1$\,K and the critical exponent $\beta = 0.335 \pm 0.01$ for the up-field sweep, and  $T_{\rm N} = 23.3 \pm 0.1$\,K and $\beta = 0.334 \pm 0.01$ for the down-field sweep. This value of the critical exponent is somewhat larger than that obtained for the AFM excitation gap at zero field [Fig.~\ref{fig:gap-Mn124}(b)] and is close to $\beta = 0.326$ for the 3D Ising model \cite{Blundell2001,Collins1989,Benner1990}. This difference can be speculatively attributed to the effect of the field polarizing the spins and by this imparting the spin system more 3D Ising character.

\begin{figure}[h]
	\centering
	\includegraphics[clip,width=0.5\columnwidth]{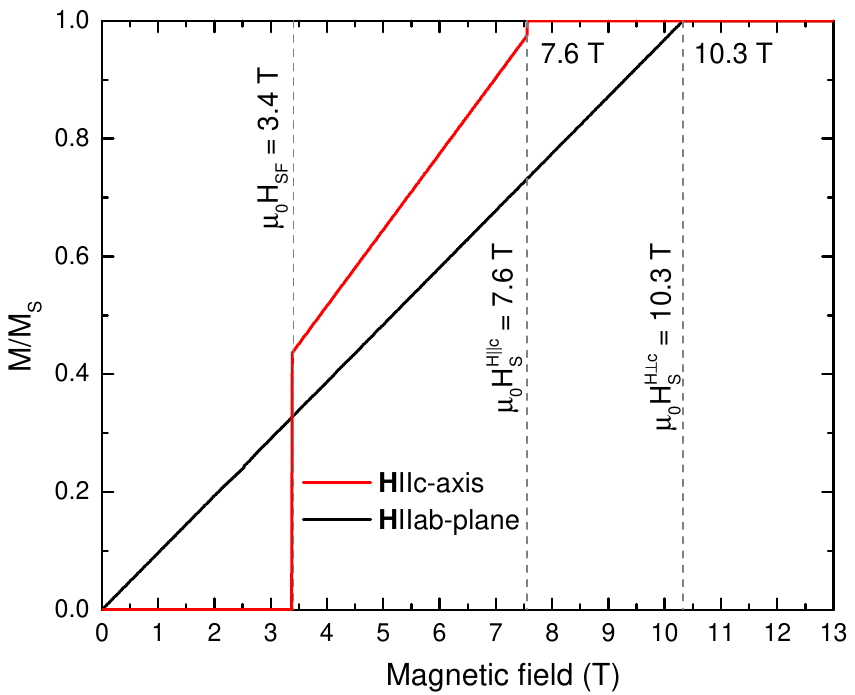}
	\caption{The calculated magnetization curves in the ordered state of \MnOTF\ for ${\bf H}\parallel c$-axis and ${\bf H}\parallel ab$-plane. Vertical dashed lines indicate the spin-flop field for ${\bf H}\parallel c$-axis and the saturation fields for the out-of-plane and in-plane orientation of the applied field.} 
	\label{calc_magn}
\end{figure}

The observed phenomenon is puzzling since no sharp anomalies in this field range were found in the static magnetization measurements for the ${\bf H}\parallel ab$-plane geometry \cite{Li2020}. 
\vk{Similar step-like changes of the microwave signal $S_D(H)$ in small fields were observed in the antiferromagnetic rare-earth ferroborate ${\rm Nd_{0.75}Ho_{0.25}Fe_{3}(BO_{3})_{4}}$ and attributed to the spin-reorientation phenomenon \cite{Kobets2015}. One can speculate that in the case of \MnOTF\ the jumps in the field range $0-1$\,T (Fig.~\ref{fig:jumps-Mn124}) can be due to the motion of the domain walls, which can cause spontaneous spin reorientation effect similar to Ref.~\cite{Kobets2015}. The even more puzzling high-field jumps could, in turn, possibly result from some spin reorientation effect due to the proximity to the AFM-FM instability revealed in our HF-ESR experiments because the step-like change of the microwave signal occurs in the field range of the crossover from the AFM to the FM spin-polarized state (Figs.~\ref{fig:modes-Mn124} and \ref{calc_magn}). 
} 
Measurements of the AC susceptibility may shed more light on the origin of this peculiar effect.

\section{Conclusions}

In summary, we have studied with HF-ESR spectroscopy the low-energy spin dynamics   in the magnetically ordered state of the single crystals of the two representatives of the \MBT\ family of magnetic topological insulators \vk{witn $n = 0$ and 1.} \MnOFS\ features an inherently ferromagnetic dynamic response of the Mn-containing septuple layers extending far above the AFM ordering temperature $T_{\rm N} = 13$\,K and by this signifying the low-dimensional nature of the spin excitations and persisting short-range FM correlations in the paramagnetic state. \MnOTF\ demonstrates at fields smaller than $\sim 6$\,T the resonance modes typical for a two-sublattice uniaxial antiferromagnet which converge to a singe paramagnetic resonance line by increasing the temperature up to the AFM ordering temperature $T_{\rm N} = 24$\,K. The energy gap for these excitations at zero field amounts to 0.45\,meV at $T = 4$\,K. The gap closes completely at $T_{\rm N}$. With the aid of the linear spin-wave theory we could quantitatively estimate all relevant parameters determining the spin dynamics in the ordered state of \MnOTF\ and \MnOFS. Notably, we found that the magnetic anisotropy of the individual septuple layers is similar in both compounds suggesting its intrinsic character little affected by the differences in the crystal structures.  

The central result of our work is the observation of a remarkable transformation of the magnetic excitations in \MnOTF\ from the AFM- to the FM-type by increasing the field strength above $\sim 6$\,T, not expected for a canonical easy-axis antiferromagnet. The FM spin dynamics in this field regime features similar properties as in \MnOFS. In particular, it evidences short-range FM correlations extending far above $T_{\rm N}$. We could rationalize this drastic field-driven change of the nature of magnetic excitations by numerically modeling the resonance modes. We have come to the conclusion that it is due to the similar scales of the inter-plane AFM exchange and the magnetic anisotropy energy of the individual septuple layers. In this situation the layers can be effectively decoupled by the applied magnetic field and demonstrate their intrinsically ferromagnetic dynamics. Our finding distinguishes \MnOTF\ as a very special compound in the family of magnetic topological insulators where the type of magnetic excitations can be tuned by a moderate magnetic field. As a possible significance of the bulk spin dynamics for the surface topological properties attracts at present significant   
attention \cite{Otrokov2019,Vidal2019,Lee2019,Li2020,Alfonsov2021}, our results on an inherently ferromagnetic dynamics of the Mn-containing layers in \MBT\ should stimulate further experimental and theoretical research along this avenue.

Finally, we observed a peculiar nonresonant effect of the step-like change (jump) of the field-dependent microwave absorption in \MnOTF\ for the in-plane field geometry. The temperature dependence of the jump field is apparently related to the magnetic order parameter but the nature of this phenomenon requires further clarification.

\section{ACKNOWLEDGMENTS}

This work was supported by the Deutsche Forschungsgemeinschaft (DFG) through grant No. KA1694/12-1 and within the Collaborative Research Center SFB 1143 ``Correlated Magnetism – From Frustration to Topology'' (project-id 247310070) and the Dresden-Würzburg Cluster of Excellence (EXC 2147) `` ct.qmat - Complexity and Topology in Quantum Matter'' (project-id 39085490). K. M. acknowledges the Hallwachs–Röntgen Postdoc Program of ct.qmat for financial support.

\bibliography{mybibfile}

\clearpage

\pagebreak

\beginappendix

\onecolumngrid

\section{Appendix}
\label{appendix}

\begin{figure}[h]
	\centering
	\includegraphics[clip,width=0.9\columnwidth]{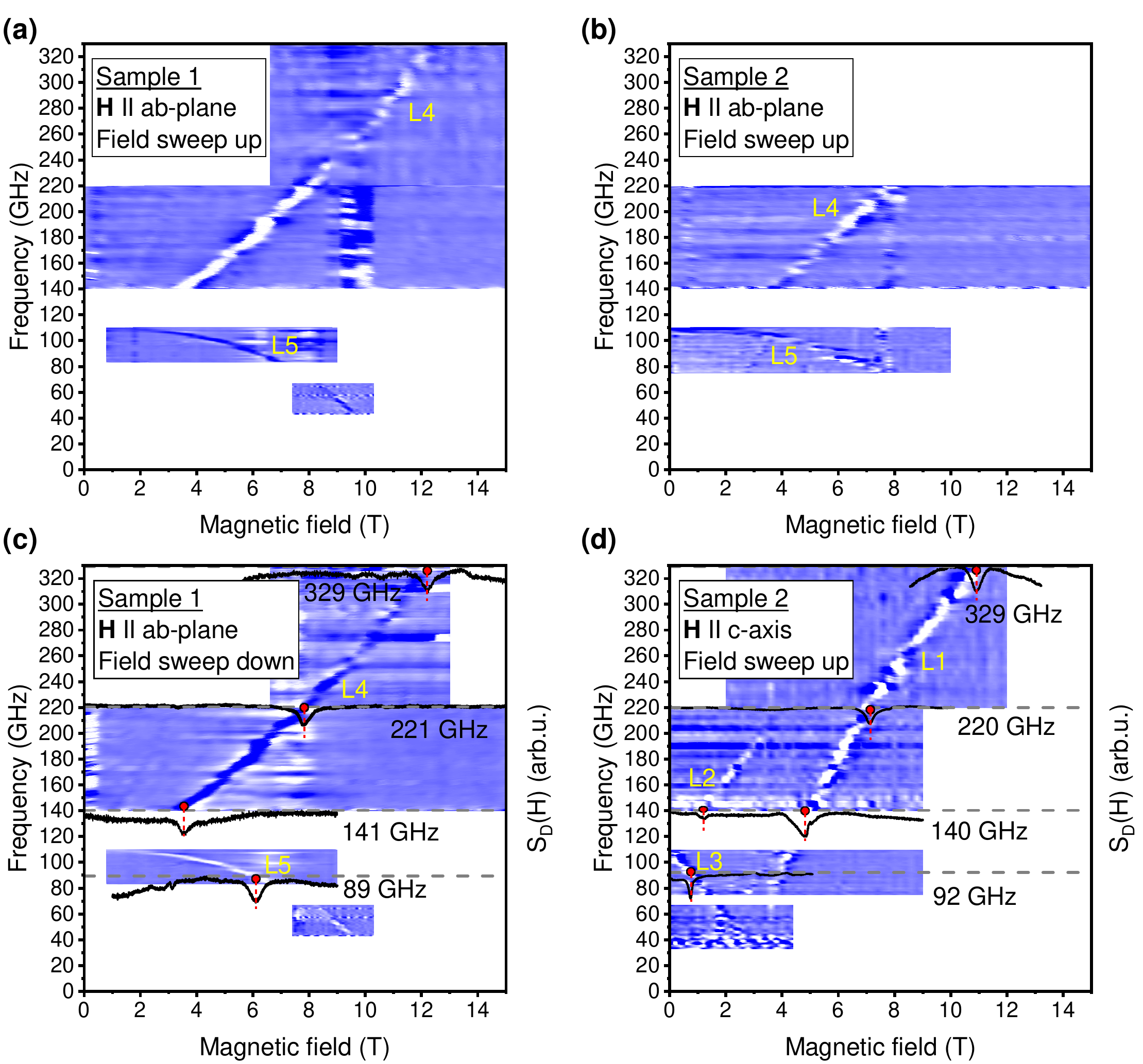}
	\caption{(a--d) Two-dimensional color maps of the resonance excitations measured on two single-crystalline samples of \MnOTF\ in the frequency-sweep mode at $T = 4$\,K. The change of the color corresponds to the relative change of the signal at the detector $S_D(\nu, H)/S_D(\nu, H-\delta H)$ by incrementing the field by $\delta H = 0.05 - 0.2$\,T. Horizontal white stripes at 0 -- 43\,GHz, 67 -- 83\,GHz and 110 -- 140\,GHz are "blind" frequency spots of the HF-ESR setup. Selected HF-ESR spectra measured in the field-sweep mode are shown as black traces in panels (c) and (d) (right vertical scale) for comparison. (a) and (c) depict results for Sample 1 measured with ${\bf H}\parallel ab$-plane by increasing the field from 0 to 15\,T or decreasing it from 15 to 0\,T, respectively; (b) and (d) depict results for Sample 2 measured by increasing the field from 0 to 15\,T applied along the $c$-axis and in the $ab$-plane, respectively.  } 
	\label{2D}
\end{figure}

Fig.~\ref{2D} presents two-dimensional color maps of the HF-ESR excitations measured on two different single-crystalline samples of \MnOTF\ at $T = 4$\,K. The measurements were performed in the frequency-sweep mode as explained in Sect.~\ref{experimental}. HF-ESR spectra at several selected frequencies recorded in the field-sweep mode are plotted there for comparison. The data obtained on different samples and in different acquisition modes demonstrate very good consistency ensuring the correctness of the $\nu(H)$ diagram presented in Fig.~\ref{fig:modes-Mn124} of the main text.

\end{document}